\documentclass[fleqn,10pt]{wlscirep}
\usepackage[utf8]{inputenc}
\usepackage[T1]{fontenc}
\title{Complex magnetic exchange, anisotropy and skyrmionic textures in two-dimensional FeXZ$_{2}$ (\textit{X} = Nb, Ta and \textit{Z} = S, Se, Te) ferromagnets} 

\author{Soheil Ershadrad}

\author{Nikola Machacova}

\author{Arnob Mukherjee}

\author{Vladislav Borisov}

\author{Olle Eriksson}

\author{Biplab Sanyal}

\affil[1]{Department of Physics and Astronomy, Uppsala University, Box-516, 75120 Uppsala, Sweden}

\affil[*]{biplab.sanyal@physics.uu.se}

\begin{abstract}
FeNbTe$_{2}$, long known as a van der Waals metallic system, has recently been resynthesized and shown to exhibit ferromagnetic order. In this study, using first-principles density functional theory (DFT), we aim to provide a deeper insight into the magnetic properties of FeNbTe$_{2}$ and its related compounds (FeXZ$_{2}$, \textit{X} = Nb, Ta; \textit{Z} = S, Se, Te), in their two-dimensional form, including their non-centrosymmetric Janus counterparts. Our results indicate that these materials are energetically, dynamically, thermally, and mechanically stable, supporting the possibility of FeNbTe$_{2}$ exfoliation and potential for experimental realization of new compounds. An evolutionary structure search suggests that FeNbTe$_{2}$ retains its monoclinic symmetry in the monolayer form. Our analysis of hopping parameters obtained from Wannierization of DFT bands shows that the nearest-neighbor magnetic interactions are primarily direct, while second-nearest and more distant interactions are mediated by the chalcogen atoms. Interestingly, although the second-nearest-neighbor interactions are smaller in magnitude, they appear to play a key role in determining the magnetic ordering in these systems. We also find evidence of canted magnetic anisotropy in FeXZ$_{2}$ compounds, with relatively strong magnetocrystalline anisotropy energy and easy-axis deviations of up to 41$^\circ$ from the out-of-plane direction—an uncommon and potentially useful feature for spintronic applications. Curie temperatures estimated from Monte Carlo simulations are below room temperature but above cryogenic levels for most compounds. Micromagnetic simulations revealed that Janus-structured FeNbSeTe can host N\'eel-type skyrmions even in the absence of an external magnetic field, making this compound a suitable candidate for further experimental studies.
\end{abstract}
\begin{document}

\flushbottom
\maketitle
%
%
\thispagestyle{empty}

\section*{Introduction}

Two-dimensional (2D) magnetic materials have attracted considerable interest in recent years, particularly following the pioneering work of exfoliation of the ferromagnetic monolayer CrI$_3$ in 2017 \cite{huang2017layer}. To date, numerous 2D magnets have been predicted and synthesized, exhibiting a wide range of electronic and magnetic properties, making this family of materials highly promising for future spintronic applications \cite{gibertini2019magnetic, gong2019two}. However, several challenges remain to be addressed. Due to their reduced thickness, 2D magnetic materials generally exhibit lower magnetic transition temperatures compared to their bulk counterparts. For instance, semiconducting 2D magnets such as CrI$_3$ and CrGeTe$_3$ have Curie temperatures (\textit{T$_{C}$}) below cryogenic levels \cite{huang2017layer, gong2017discovery}. In contrast, metallic 2D magnets can achieve relatively higher transition temperatures, due to the long-range exchange interactions facilitated by conduction electrons. Recent advancements, including the synthesis and exfoliation of few-layer metallic Fe$_5$GeTe$_2$ with a \textit{T$_{C}$} of 280 K \cite{may2019ferromagnetism, may2019physical}, and the further enhancement of \textit{T$_{C}$} through doping in this material \cite{ghosh2024structural, ngaloy2024strong}, demonstrate the potential of 2D metallic magnets to achieve magnetism above room temperature. Recent studies have reported the successful application of room-temperature 2D metallic magnets in spin-valve devices and spin-orbit torque (SOT) systems \cite{zhao2023room, zhao2023coexistence}. Furthermore, there is a growing demand for 2D magnets with enhanced magnetic properties, such as strong magnetic anisotropy energy (MAE) \cite{ershadrad2024ab}. For example, despite the promising \textit{T$_{C}$} of Fe$_5$GeTe$_2$, these compounds exhibit relatively low magnetic anisotropy energy (MAE), on the order of a few $\mu$eV per Fe atom \cite{ershadrad2022unusual}. The prediction of new materials using ab initio techniques provides an efficient and rapid approach for screening promising 2D materials. Many synthesized 2D materials were initially predicted through such methods \cite{lebegue2013two, mounet2018two}.

Skyrmions represent another emerging magnetic feature with significant potential for applications as information carriers. These are topologically protected magnetic textures that arise in non-centrosymmetric magnets due to Dzyaloshinskii-Moriya interactions (DMI) \cite{dzyaloshinsky1958thermodynamic, moriya1960anisotropic}. The DMI is an asymmetric exchange interaction originating from spin–orbit coupling in magnetic systems lacking inversion symmetry. This interaction aligns spins in twisted pairs, leading to the formation of various chiral magnetic structures, such as skyrmions \cite{yang2023first}. DMI can also stabilize non-topological non-collinear states like spin spirals. 

FeXZ$_{2}$ (\textit{X} = Nb, Ta and \textit{Z} = S, Se, Te) represents a family of metallic van der Waals (vdW) magnets that can be exfoliated down to the monolayer regime. The layered crystal FeNbTe$_{2}$, in its low-temperature (LT) orthorhombic phase, was first synthesized over 30 years ago \cite{li1992new}. Experimental results suggested that this phase behaves as an Anderson insulator with spin glass characteristics \cite{zhang1997magnetic}. In addition to the LT phase, a high-temperature (HT) phase was identified, which crystallizes in a monoclinic structure and exhibits a clear ferromagnetic order with a transition temperature of approximately 70 K \cite{wu2024tailoring, stepanova2024bulk}. The LT phase can be transformed into the HT phase through simple thermal annealing.

A similar phase transition has been reported in NiNbTe$_{2}$, although no magnetic ordering was observed in this compound \cite{neu2019orthorhombic}. Other members of this family, such as CoNbTe$_{2}$ and CoTaTe$_{2}$, were found to exhibit both monoclinic and orthorhombic symmetries. However, these compounds were reported to display a Pauli paramagnetic phase \cite{jing1993x, li1992new, tremel1991isolated}.

In the HT (monoclinic) phase of FeNbTe$_{2}$, metallic behavior has been observed, along with a large negative magnetoresistance (MR) across the entire temperature range and an anomalous Hall effect (AHE) below the ferromagnetic transition temperature \cite{bai2019intrinsic, qi2022abnormal, wu2024tailoring}. Additionally, FeNbTe$_{2}$ demonstrates directional elastic anisotropy and exhibits promising optical absorption in the ultraviolet (UV) region \cite{aktary2023mechanical}.

Several experimental and theoretical studies have been conducted on certain FeXZ$_{2}$ compounds in their bulk forms. However, to the best of our knowledge, a comprehensive investigation of these compounds in the monolayer regime remains unavailable. In this work, we address this gap by systematically exploring the structural, electronic, and magnetic properties of FeXZ$_{2}$ (\textit{X} = Nb, Ta and \textit{Z} = S, Se, Te) monolayers in their monoclinic phase, using \textit{ab initio} methods coupled to Monte Carlo simulations as well as micromagnetic simulations. Our evolutionary structural search calculations indicate that the monoclinic phase is energetically favorable even in the monolayer regime. We aim to provide insights into the exchange mechanism and magnetic anisotropy energy of these compounds and explore the potential for skyrmion formation in their Janus counterparts.

\section*{Results and Discussion}
\subsection*{Structural properties and stability}
\begin{figure*}[htp]
 \centering
 \includegraphics[width=\linewidth]{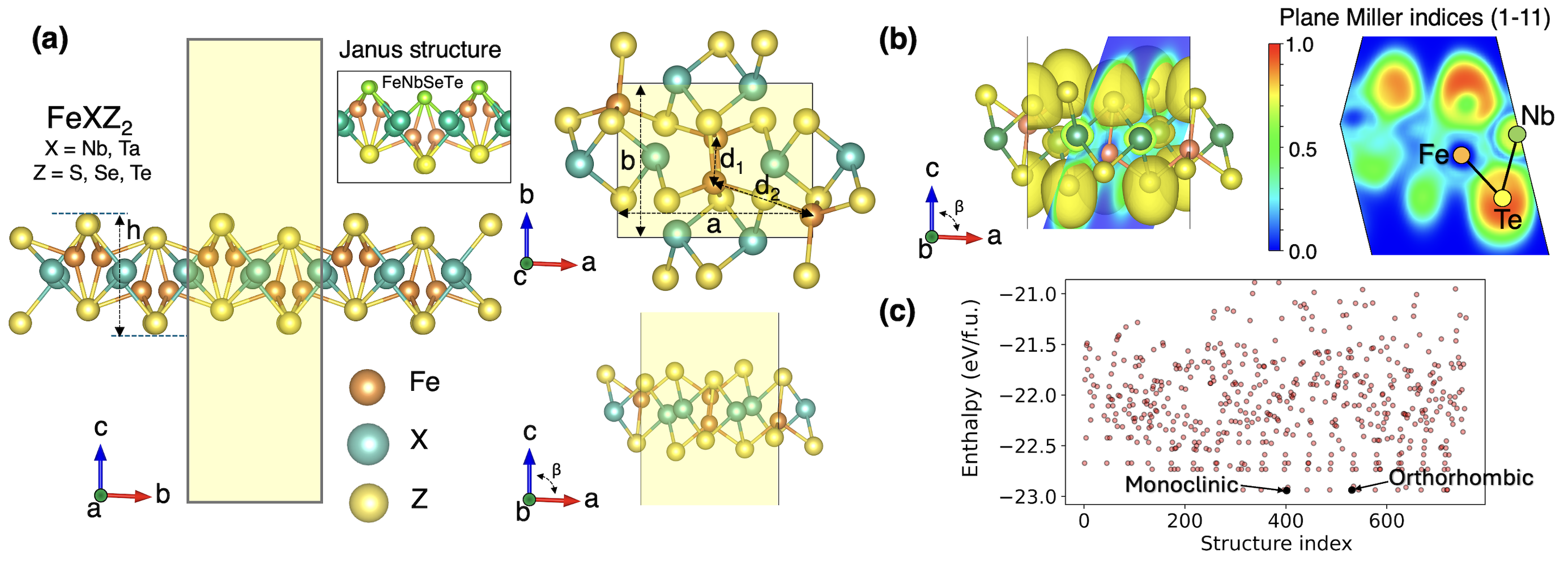}
  \caption{(a) Structural configuration of the FeXZ$_{2}$ monolayer depicted from both top and side views. The yellow-shaded region highlights the unit cell, where $a$ and $b$ denote the lattice parameters, $h$ represents the thickness, $d_{1}$ and $d_{2}$ correspond to the first and second nearest-neighbor Fe-Fe distances, respectively, and $\beta$ indicates the monoclinic angle. (b) Electron localization function (ELF) for FeNbTe${2}$, illustrated on the (1-11) plane as a 2D isosurface. The circles schematically represent the positions of Fe, Nb, and Te atoms. (c) The enthalpies of the allotropes found in evolutionary structure search for FeNbTe$_{2}$ monolayer. The energies of monoclinic and orthorhombic structures are pointed by arrows.}
  \label{fig1}
\end{figure*}

We investigated the structural properties of the high-temperature phase of FeXZ$_{2}$ systems, as they host magnetic order according to the experimental results. Fig. \ref{fig1}(a) illustrates the schematic structural properties of the FeXZ$_{2}$ monolayer, where the yellow-shaded area represents the unit cell. The unit cell consists of four Fe atoms, four \textit{X} atoms and eight \textit{Z} atoms (i.e., Fe$_{4}$X$_{4}$Z$_{8}$). The optimized lattice parameters are summarized in Table \ref{tab1}. Consistent with experimental observations, all structures exhibit the monoclinic symmetry with a space group of P2$_{1}$/c (No. 14) with inversion symmetry (except for Janus systems). For the FeNbTe$_{2}$ monolayer, we determined lattice parameters of $a = 8.04$ \AA, $b = 6.34$ \AA, and a monoclinic angle of $\beta = 92.32^\circ$, which are in good agreement with the experimental bulk values of $a = 7.93$ \AA, $b = 6.25$ \AA, and $\beta = 92.11^\circ$ \cite{wu2024tailoring}. 
As expected, the lattice parameters increase along the chalcogen group; however, the monoclinic angle ($\beta$) remains nearly constant at approximately 92.3$^\circ$ across all systems. In the Ta-hosted system, a reduction in lattice parameters is observed compared to the Nb-based counterparts. This reduction is partially compensated by a slight increase in the thickness of the monolayer. The shrinkage in the lattice parameters may be attributed to the lanthanide contraction effect, which arises from the poor shielding of the 4$f$ electrons in the Ta atoms.

Fig. \ref{fig1}(b) presents the electron localization function (ELF) as a 3D (top) and a 2D isosurface along the (1-11) plane (bottom) for FeNbTe$_{2}$. The ELF reveals that the electrons are localized around the chalcogen and Nb atoms, while a region of delocalized electrons is observed around the Fe atoms. Generally, localized electrons are indicative of covalent bonding; thus, it can be inferred that the Nb-Te bonds exhibit a covalent character. In contrast, the absence of electron localization along the Fe-Fe pair ($d = 2.4$ \AA) suggests a metallic bonding nature. Accordingly, the crystal structure can be described as Fe pairs embedded within a covalently bonded Nb-Te skeleton.
Further Bader charge analysis (see Table S1 in the supplemental materials) reveals charge depletion around Fe and Nb(Ta) atoms, accompanied by charge accumulation around the chalcogen atoms. The magnitude of charge transfer is proportional to the electronegativity of the chalcogen species, with S, Se, and Te atoms gaining approximately 1, 0.75, and 0.5 $e$, respectively.


\begin{table*}[t]
\centering
\small
\setlength{\tabcolsep}{3pt}
\renewcommand{\arraystretch}{1.5}
\caption{The optimized lattice constants, $a$, $b$ (\AA); monoclinic angle, $\beta$ ($^\circ$); thickness, $h$ (\AA); first and second nearest neighbor Fe-Fe distance, $d_{i}$ (\AA); cohesive energy per atom, $E_{coh}$ (eV/atom); formation energy per atom, $E_{form}$ (eV/atom); elastic constants, $C_{ij}$ (GPa); for FeXZ$_{2}$ monolayers.} 
\begin{tabular}{l|c|c|c|c|c|c|c|c|c|c|c|c}
\hline\hline 
Structure         & $a$      & $b$      & $\beta$   & $h$      & $d_{1}$ & $d_{2}$ & $E_{coh}$ & $E_{form}$  & $C_{11}$ & $C_{12}$ & $C_{22}$ & $C_{44}$ \\ 
                  & (\AA)    & (\AA)    & ($^\circ$) & (\AA)    & (\AA)   & (\AA)   & (eV/atom) & (eV/atom)   & (GPa)    & (GPa)    & (GPa)    & (GPa)    \\
\hline 
FeNbS$_2$    & 7.80 & 5.98 & 92.39   & 3.85 & 2.42 & 4.13  & 5.70 & -0.69  & 115.0  & 22.2 & 101.9 & 34.7  \\
FeNbSe$_2$   & 7.91 & 6.14 & 92.35   & 4.18 & 2.38 & 4.18  & 5.27 & -0.57  & 105.3  & 17.6 & 90.6  & 33.1  \\
FeNbTe$_2$   & 8.04 & 6.34 & 92.32   & 4.62 & 2.39 & 4.24  & 4.86 & -0.33  & 89.0   & 10.6 & 81.9  & 30.3  \\
FeNbSeTe     & 7.98 & 6.24 & 92.33   & 4.39 & 2.39 & 4.21  & 5.06 & -0.45  & 98.4   & 13.9 & 86.0  & 31.6  \\ 
FeTaS$_2$    & 7.74 & 5.96 & 92.41   & 3.85 & 2.38 & 4.09  & 5.95 & -0.66  & 113.4  & 23.2 & 104.2 & 36.2  \\
FeTaSe$_2$   & 7.78 & 6.09 & 92.39   & 4.21 & 2.35 & 4.09  & 5.51 & -0.52  & 110.4  & 17.9 & 92.1  & 33.3  \\
FeTaTe$_2$   & 7.92 & 6.28 & 92.35   & 4.63 & 2.37 & 4.16  & 5.08 & -0.28  & 103.9  & 10.6 & 81.9  & 30.3  \\
FeTaSeTe     & 7.84 & 6.19 & 92.37   & 4.42 & 2.37 & 4.11  & 5.30 & -0.40  & 99.5  & 12.0 & 86.1  & 32.9  \\
\hline\hline 
\end{tabular}
\label{tab1}
\end{table*}

\begin{table*}[]
\normalsize
\setlength{\tabcolsep}{3pt}
\renewcommand{\arraystretch}{1.5}
\caption{Magnetic properties of FeXZ$_{2}$ monolayers: Relative total energies of various antiferromagnetic (AFM) configurations with respect to the ferromagnetic (FM) ground state (eV/Fe); averaged spin magnetic moments per Fe (\( M^{\text{spin}}_{\text{Fe}} \)), Nb(Ta) (\( M^{\text{spin}}_{X} \)), and chalcogens (\( M^{\text{spin}}_{Z} \)) (\(\mu_B\)); averaged orbital magnetic moments per Fe (\( M^{\text{orb.}}_{\text{Fe}} \)) (\(\mu_B\)); spin polarization at the Fermi level (\(\%\)); exchange stiffness constant (\(\text{meV} \cdot \text{\AA}^{2}\)); spiralization value (\(\text{meV} \cdot \text{\AA}\)); magnetocrystalline anisotropy energy (MAE) (meV/Fe), defined as energy difference between hard and easy-axis; and magnetic transition temperature (\textit{T$_{C}$}) (K).} 
\small
\begin{tabular}{l|c|c|c|c|c|c|c|c|c|c|c|c}
\hline\hline 
Structure  & AFM$_{1}$  & AFM$_{2}$  & AFM$_{3}$     & M$^{spin}_{Fe}$ & M$^{spin}_{X}$ & M$^{spin}_{Z}$ & M$^{orb.}_{Fe}$ & Spin pol. & A & D & MAE  & T$_{C}$  \\

      & (eV/Fe)     & (eV/Fe)    & (eV/Fe)      & ($\mu_B$)   & ($\mu_B$)  & ($\mu_B$)   & ($\mu_B$) &  (\%) & (meV.\AA$^{2}$)  & (meV.\AA) & (meV/Fe)  & (K) \\

\hline 
FeNbS$_2$    & 0.092 & 0.116 & 0.187 & 2.20 & -0.32 & 0.07 & 0.061 & -0.7 & 302.8 & 0.0 & 0.28 & 177 \\
FeNbSe$_2$   & 0.091 & 0.133 & 0.219 & 2.21 & -0.39 & 0.06 & 0.077 & -3.3 & 337.9 & 0.0 & 0.46 & 199 \\
FeNbTe$_2$   & 0.035 & 0.129 & 0.172 & 2.09 & -0.27 & 0.06 & 0.096 & -0.5 & 313.3  & 0.0 & 0.47 & 115  \\
FeNbSeTe & 0.052 & 0.087 & - & 2.39 (1.86) & -0.28 (-0.34) & 0.05 & 0.087 & 5.6 & 259.7  & 20.4 & 0.41 & 139  \\ 
FeTaS$_2$  & 0.087 & 0.121 & 0.190 & 2.10 & -0.24 & 0.07 & 0.094 & -14.5 & 421.7 & 0.0 & 1.01 & 195  \\
FeTaSe$_2$ & 0.039 & 0.136 & 0.190 & 1.90 & -0.39 & 0.04 & 0.109 & -47.2 & 232.7 & 0.0 & 1.57 & 155  \\
FeTaTe$_2$ & 0.014 & 0.127 & 0.151 & 1.74 & -0.29 & 0.01 & 0.122 & -48.2 & 169.3   & 0.0 & 1.26 & 99 \\
FeTaSeTe & 0.012 & - & - & 2.07 (1.50) & -0.34 (-0.31) & 0.03 & 0.115  & -36.7  &  157.3  & 13.5 & 1.34 & 133  \\
\hline\hline 
\end{tabular}
\label{tab2}
\end{table*}

To ensure the energetic stability of all FeXZ$_{2}$ monolayers, we calculated their cohesive energy ($E_{coh}$) and formation energy ($E_{form}$) using the following expressions:

\begin{equation}
E_{coh} = \frac{4E_{Fe} + 4E_{X} + 8E_{Z} - E_{T}}{16}	
\end{equation}

\begin{equation}
E_{form} = \frac{E_{T} - 4E_{Fe} - 4E_{X} - 8E_{Z}}{16}	
\end{equation}
Here, $E_{T}$ represents the total energy of the crystalline system, while $E_{Fe}$, $E_{X}$, and $E_{Z}$ denote the total energies of a single free-standing atom of Fe, Nb(Ta), and S(Se,Te) for cohesive energy calculations, and their bulk forms for formation energy calculations. For each element, the most stable bulk form was obtained from the Materials Project database~\cite{jain2013commentary} and re-optimized using consistent computational parameters. The computed values are listed in Table~\ref{tab1}. 

Cohesive energy represents the energy required to decompose a crystalline structure into its constituent atoms, serving as a measure of the binding energy between atoms. According to Eq.~(1), a positive $E_{coh}$ indicates the system is energetically stable. All FeXZ$_{2}$ monolayers exhibit positive cohesive energy, confirming their cohesive stability. The magnitude of $E_{coh}$ increases with the electronegativity of the chalcogen atoms. Among the compounds, FeNbTe$_{2}$ shows the lowest cohesive energy of 4.86~eV/atom, which is still significantly higher than that of CrI$_{3}$ (2.32~eV/atom)~\cite{chen2022alloy}. 

Formation energy, the second criterion for energetic stability, is calculated relative to the most stable bulk phase of each constituent element. A negative $E_{form}$ indicates that a compound is stable against decomposition into the bulk forms of its elemental constituents. As shown in Table~\ref{tab1}, all FeXZ$_{2}$ monolayers exhibit negative formation energies, demonstrating their energetic favorability. Similar to cohesive energy, the magnitude of $E_{form}$ is also proportional to the electronegativity of the chalcogen atoms. FeNbTe$_{2}$ and FeTaTe$_{2}$ monolayers have the least favorable $E_{form}$, with -0.33 and -0.28 eV/atom, respectively. These values can be compared to that of CrI$_{3}$ (-0.18~eV/atom)~\cite{mcguire2015coupling} and Fe$_{3}$GeTe$_{2}$ (-0.08~eV/atom)~\cite{liu2022layer}. 
Although the calculated cohesive and formation energies cannot guarantee the synthesizability of a compound per se, favorable cohesive and formation energies serve as promising indicators for the likelihood of successful synthesis in FeXZ$_{2}$ monolayers.

Building on the confirmation of their energetic stability, we further investigated the dynamical (vibrational) stability of the FeXZ$_{2}$ compounds by analyzing their phonon spectra, as shown in Fig. S1 in the supplemental materials. The phonon spectra reveal the absence of imaginary frequencies, signifying structural stability. Within the framework of the harmonic approximation, imaginary frequencies are indicative of repulsive interatomic forces that render a structure unstable. 
In three-dimensional systems, atoms possess three degrees of freedom for collective motion along the $x$, $y$, and $z$-axes. For FeXZ$_{2}$ unit cells, which comprise 16 atoms, this results in 48 vibrational modes. Of these, three modes are acoustic (ZA, TA, and LA) with zero frequency at the $\Gamma$ point, while the remaining 45 modes are optical. Unlike conventional three-dimensional crystals, where all acoustic modes exhibit linear dispersion near the $\Gamma$ point, two-dimensional materials are characterized by a quadratic dispersion in the out-of-plane acoustic branch (ZA) \cite{katsnelson2012physics}. 

To examine the thermal behavior of FeXZ$_{2}$ monolayers at an elevated temperature, we conducted ab initio molecular dynamics (AIMD) simulations for a duration of 6 ps. The energy variations throughout the simulation, along with the final atomic configuration after 6 ps at $T = 500$~K, are depicted in Fig. S2 in the supplemental materials. The energy variations showed minor fluctuations, remaining close to their average values, thereby confirming the thermal stability of all structures under elevated temperature conditions. Structural snapshots from the simulation reveal no evidence of broken chemical bonds, ensuring the preservation of the overall structural integrity of the FeXZ$_{2}$ monolayers. These results indicate that FeXZ$_{2}$ compounds retain their thermal stability above room temperature(at 500~K).

Strain energy density (W) represents the energy stored per unit volume in a material due to deformation. To have mechanical stability in crystals, the strain energy density,
\begin{equation}
W = \frac{1}{2} \boldsymbol{\epsilon}^T \boldsymbol{C} \boldsymbol{\epsilon}
\end{equation}
needs to remain positive for any nonzero strain $\epsilon$ \cite{born1954dynamical,mouhat2014necessary}. This implies that the stiffness matrix $\boldsymbol{C}$ must be positive definite. A matrix is positive definite if and only if all its eigenvalues are greater than zero. 
Symmetry constraints dictate the following form for the stiffness matrix in a monoclinic system:

\[
\begin{bmatrix}
C_{11} & C_{12} & C_{13} & 0      & 0      & C_{16} \\
C_{12} & C_{22} & C_{23} & 0      & 0      & C_{26} \\
C_{13} & C_{23} & C_{33} & 0      & 0      & C_{36} \\
0      & 0      & 0      & C_{44} & C_{45} & 0      \\
0      & 0      & 0      & C_{54} & C_{55} & 0      \\
C_{16} & C_{26} & C_{36} & 0      & 0      & C_{66}
\end{bmatrix}
\]
However, in the case of monolayers, all elements with $z$-components are effectively zero. Therefore, the effective stiffness matrix can be reduced to:
\[
\begin{bmatrix}
C_{11} & C_{12} & 0  \\
C_{12} & C_{22} & 0  \\
0      & 0      & C_{44}
\end{bmatrix}
\]
The reduced stiffness matrix for FeXZ$_{2}$ monolayers, along with their corresponding eigenvalues obtained after diagonalization, are presented in Fig. S3 of the supplementary materials. It can be noted that all structures comply with the mechanical stability criteria.

We further employed an evolutionary structure search algorithm to explore the energy landscape of the FeNbTe$_{2}$ monolayer. More than 600 two-dimensional structures were generated using evolutionary algorithms, incorporating heredity, mutation, and randomness to identify configurations with the lowest enthalpy. Fig.~\ref{fig1}(c) presents the enthalpy vs. structural number, where the monoclinic and orthorhombic phases are highlighted. It should be noted that during the cell optimization steps, several parent structures can converge into a single structure represented in different unit cell symmetries, resulting in structural indices with nearly identical enthalpies. While the monoclinic and orthorhombic phases are more energetically favorable than any other symmetry, they exhibit a very small energy difference relative to each other, with the monoclinic phase being only 0.17 meV/f.u. lower in energy. These findings confirm that the monoclinic phase remains energetically favorable for FeNbTe$_{2}$ in the monolayer regime.

\subsection*{Magnetic properties}

\subsubsection*{Magnetic ground state and magnetic moments}
To determine the magnetic ground state of FeXZ\(_{2}\) monolayers, four distinct magnetic configurations were evaluated: the ferromagnetic (FM) order and three symmetrically nonequivalent antiferromagnetic (AFM) orders, labeled AFM\(_{1}\), AFM\(_{2}\), and AFM\(_{3}\). These configurations are schematically illustrated in Fig. S4 in the supplemental materials. It was found that the FM order exhibits the most favorable energy in all systems. The relative energies of the AFM states with respect to the FM state are presented in Table~\ref{tab2}. The missing values correspond to magnetic orders that could not be stabilized, leading to a spontaneous transition to other magnetic orders. It is noteworthy that AFM\(_{1}\) is closer in energy to the FM order compared to AFM\(_{2}\) and AFM\(_{3}\). This indicates that antiparallel spin alignment on Fe-Fe pairs (with a bond length of approximately \(d \sim 2.4 \, \text{\AA}\)) is not energetically favorable. In contrast, neighboring pairs can align antiparallel to each other with a minimal energy cost. The obtained FM ground state aligns well with experimental results reported for synthesized FeNbTe\(_{2}\) \cite{wu2024tailoring}. 
The averaged spin and orbital moments on atomic sites are listed in in Table~\ref{tab2}. While the induced spin moments on chalcogen atoms are very small, Nb and Ta carry an anti-parallel moment of approximately -0.3~$\mu_B$. The orbital moments on Fe sites exceed 0.06~$\mu_B$, increasing in the vicinity of elements with higher spin-orbit coupling up to 0.12~$\mu_B$. The orbital moments of other elements were negligible.

We have also estimated the spin polarization of all the systems using the expression $SP=\frac{n(E_F)^{\uparrow}-n(E_F)^{\downarrow}}{n(E_F)^{\uparrow}+n(E_F)^{\downarrow}}$, displayed in Table 2. $n(E_F)$ is the density of states (DOS) at the Fermi level $E_F$ for $\uparrow/\downarrow$ spin extracted from the spin-polarized DOS plots shown in the Supplementary Information. One can observe a variation of positive and negative spin polarizations. The highest value is obtained for FeTaTe$_2$ with a negative spin polarization.

\subsubsection*{Magnetic exchange interactions and exchange mechanism}

Isotropic Heisenberg exchange interaction  (\(J_{ij}\)) and Dzyaloshinskii-Moriya interactions (DMI) (\(D_{ij}\)) parameters as a function of pair distances are calculated and reported in Fig.~\ref{fig2}(a) for FeNbZ\(_{2}\) monolayers, and Fig. S7-S14 in the supplemental materials for FeTaZ\(_{2}\) monolayers. It can be noted that the strongest exchange interaction occurs between the first nearest-neighbor pair (i.e., Fe1 and Fe2) for all systems. \(J_{12}\) is strongly ferromagnetic, varying between 82.6 meV (FeNbS\(_2\)) and 104.8 meV (FeNbSe\(_2\)). The second nearest-neighbor interaction (\(J_{13}\)) is also the second strongest interaction, varying between 25.7 meV (FeTaS\(_2\)) and 9.1 meV (FeTaTe\(_2\)). The third nearest-neighbor interaction (second blue square) is antiferromagnetic (AFM) in most systems, but its strength is very small and cannot overcome the strong ferromagnetic (FM) interactions.

In FeNbS\(_2\), the strongest DMI is \(D_{11} = 0.3\) meV. While \(D_{11}\) remains close to 0.3 meV across all Nb-based systems, the magnitudes of \(D_{14}\) and \(D_{13}\) increase in compounds with stronger chalcogen spin-orbit coupling, i.e. FeNbSe\(_2\) (by approximately a factor of two) and FeNbTe\(_2\) (by approximately a factor of four). This enhancement can be attributed to indirect exchange mediated by chalcogen atoms in \(D_{13}\) and \(D_{14}\), a mechanism that will be discussed in detail later. In Ta-based systems, where Ta serves as an additional source of strong SOC, \(D_{14}\) and \(D_{13}\) exhibit the highest magnitudes.

In centrosymmetric structures, the DMI for centrosymmetric pairs, such as \( D_{12} \), is inherently zero. Additionally, the DMI contributions from non-centrosymmetric pairs cancel each other out, leading to a vanishing spiralization value. However, in Janus systems with broken inversion symmetry (e.g., FeNbSeTe and FeTaSeTe), \(D_{12}\) becomes the dominant DMI component, reaching 0.8 meV in FeNbSeTe and 1.2 meV in FeTaSeTe. The asymmetry in the DMI of these systems prevents complete cancellation, resulting in a finite spiralization value, which will be discussed in the following sections.

To explore the exchange mechanism, we computed the orbital-decomposed exchange interactions within the FeXZ\(_2\) family. The results for the first and second nearest neighbors interactions of FeNbTe\(_2\) are presented in Fig.~\ref{fig2}(b and c), while the corresponding data for other compounds are provided in Fig. S15 and S16, in the supplementary materials. A similar behavior is observed in the orbital-decomposed exchange interactions of all compounds. For \( J^{1\text{st}} \) (Fe\(_{12}\)), the dominant contribution arises from the \( d_{yz}\)-\( d_{yz} \) interaction (20.4 meV). Examining the schematic exchange path of this interaction and the corresponding hopping parameters obtained from Wannierization of DFT bands, represented by red double-headed arrows in units of meV (see Fig.~\ref{fig2}(b)), reveals an overlap between the \( d_{yz} \) orbitals of Fe\(_1\) and Fe\(_2\), leading to a strong hopping parameter of 50 meV. This indicates that the mechanism for the first-neighbor interactions is direct exchange.

In \( J^{2\text{nd}} \) (Fe\(_{13}\)), no significantly strong orbital-decomposed exchange interaction is identified. The strongest contributions, highlighted in orange, are approximately an order of magnitude smaller than the first-neighbor \( d_{yz}\)-\( d_{yz} \) interaction. The second-nearest-neighbor pairs are 4.24~\AA\ apart, and for both \( d_{z^2}\)-\( d_{z^2} \) and \( d_{xy}\)-\( d_{x^2-y^2} \) interactions, no direct overlap is structurally possible, as inferred from the zero hopping parameters between these orbitals. Therefore, only an indirect exchange mechanism is feasible for \( J^{2\text{nd}} \). Extracted hopping parameters suggest that the indirect exchange is mediated via \( p \)-orbitals of the chalcogen atoms. Specifically, for the \( d_{z^2}\)-\( d_{z^2} \) and \( d_{xy}\)-\( d_{x^2-y^2} \) interactions, the dominant mediating agents are \( p_y \) and \( p_z \), respectively, as shown in Fig.~\ref{fig2}(c). The hopping parameters along these indirect exchange paths are relatively strong (\( t_{ij} > 40 \) meV). Moreover, more complex indirect exchange pathways involving both chalcogens and Nb (or Ta) can be identified for third-nearest-neighbor pairs and beyond.

 \begin{figure*}[t!]
 \centering
 \includegraphics[width=\linewidth]{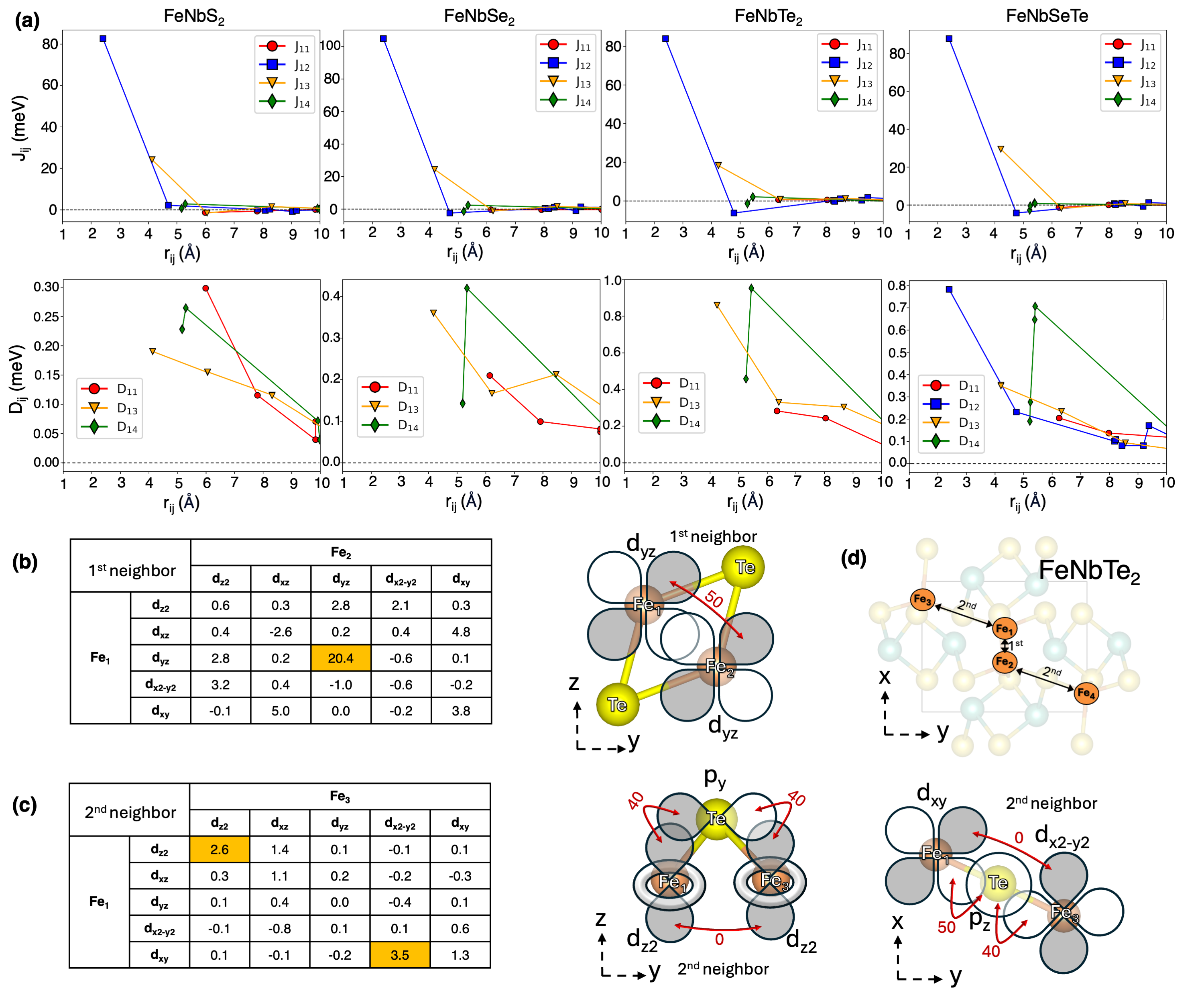}
  \caption{(a) Isotropic Heisenberg exchange interactions (\(J_{ij}\)) (top panels) and Dzyaloshinskii-Moriya interactions (\(D_{ij}\)) (bottom panels) as a function of pair distance, expressed in meV, for FeNbZ\(_2\) monolayers. For visual clarity, only interactions with \(i = 1\) are shown (see Supplemental Materials for FeTaZ\(_2\) and exchange interactions with \(i \neq 1\)). (b) and (c) Orbital-decomposed exchange interactions in FeNbTe\(_2\) and schematics of exchange pathways for the first (Fe\(_{12}\)) and second (Fe\(_{13}\)) nearest-neighbor interactions, respectively. The strongest decomposed interactions are highlighted in orange. The schematics include the involved \( p \)- and \( d \)-orbitals, depicted in gray and white, while the red double-headed arrows represent the Wannierization hopping parameters between orbitals, given in units of meV. (d) Top view of the unit cell, illustrating the first and second nearest-neighbor Fe pairs.
}
  \label{fig2}
\end{figure*}

\subsubsection*{Magnetocrystalline anisotropy Energy (MAE)}
In general, MAE of a crystal in the lowest symmetry (i.e. triclinic system) can be expressed as \cite{blundell2001magnetism},

\begin{equation}\label{eq1_mae}\footnotesize
E = K_{0} + K_{1} \alpha_{1}^{2} + K_{2} \alpha_{2}^{2} + K_{3} \alpha_{3}^{2} + K_{4} \alpha_{1} \alpha_{2} + K_{5} \alpha_{2} \alpha_{3} + K_{6} \alpha_{3} \alpha_{1}
\end{equation}
where $E$ is the total energy, $K_{i}$s are first-order magnetic anisotropy constants, $\alpha_{1} = \sin \theta \cos \phi~,~\alpha_{2} = \sin \theta \sin \phi~and~\alpha_{3} = \cos \theta$, where $\theta$ and $\phi$ are Euler angles.

\begin{figure}[t!]
 \centering
 \includegraphics[width=\linewidth]{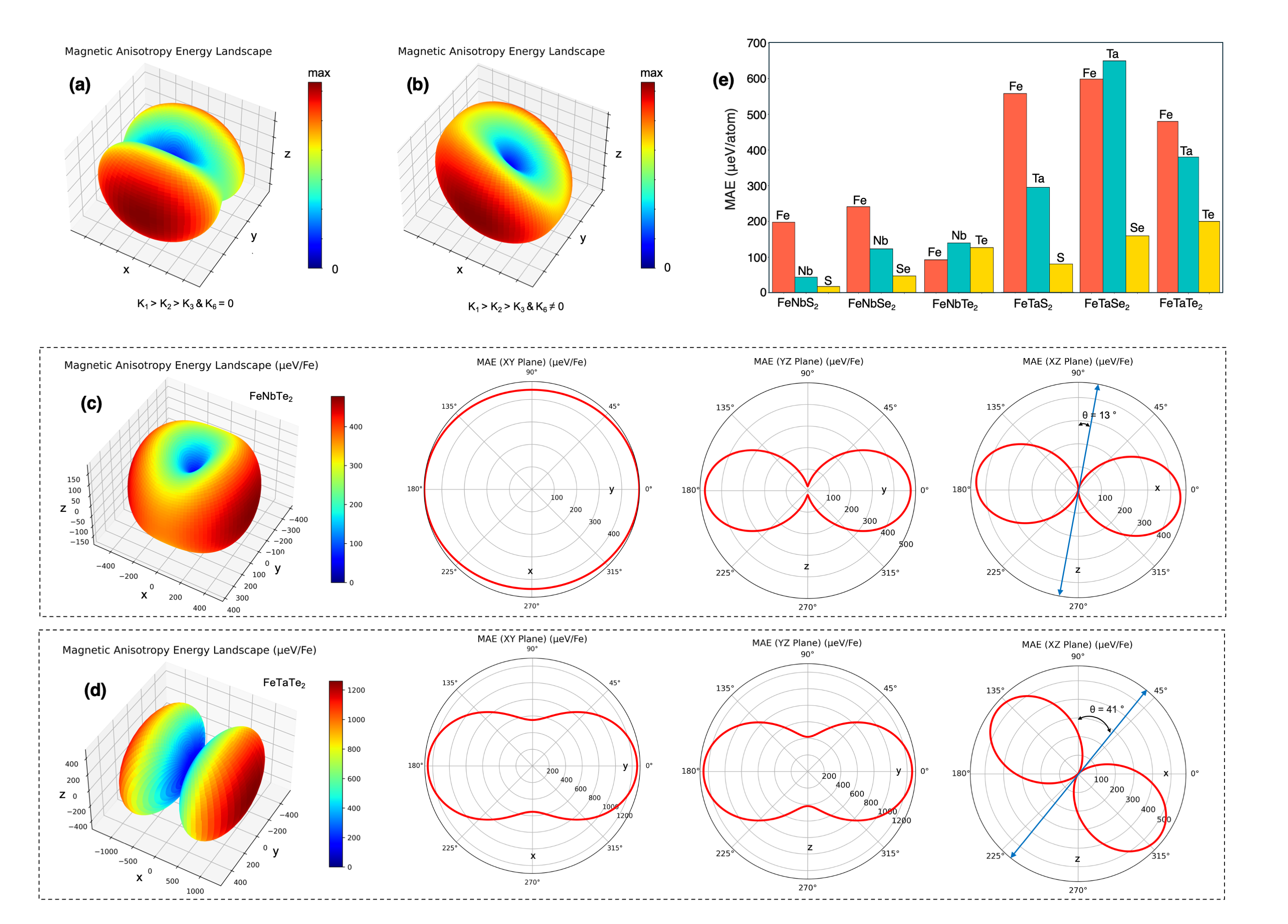}
  \caption{Three-dimensional (3D) and projected two-dimensional (2D) magnetic anisotropy energy (MAE) surfaces, expressed in units of \(\mu\)eV/Fe:  
(a) and (b) Schematic representations of generic 3D MAE surfaces, illustrating the influence of first-order magnetic anisotropy constants (\(K_i\)) on the energy landscape.  
(c) and (d) Computed 3D MAE surfaces for FeNbTe\(_2\) and FeTaTe\(_2\), respectively.  
(e) and (f) Projected 2D MAE surfaces for FeNbTe\(_2\) and FeTaTe\(_2\), respectively, onto the \(xy\)-, \(yz\)-, and \(xz\)-planes, where the tilted angle, denoted by \(\theta\), is indicated by a blue double-headed arrow. (f) Atom projected magnetic anisotropy energy for FeXZ\(_2\) monolayers in units of $\mu$eV/atom, where orange, turquoise, and yellow bars depict Fe, Nb(Ta), and chalcogen atoms.
}
  \label{fig3}
\end{figure}

Considering the monoclinic symmetry in all FeXZ$_{2}$ compounds, Eq. \ref{eq1_mae} simplifies to,

\begin{equation}\label{eq2_mae}\small
E = K_{0} + K_{1} \alpha_{1}^{2} + K_{2} \alpha_{2}^{2} + K_{3} \alpha_{3}^{2} + K_{6} \alpha_{1} \alpha_{2}
\end{equation}
Eq. \ref{eq2_mae} shows that, when $K_{6}$ is small, it is the relative ratio between $K_{1-3}$ that determines the three-dimensional surface for MAE, where six permutation scenarios are possible $\{K_{i} < K_{j} < K_{k}\}$. The generic form of MAE surface for a scenario where ${K_{3} < K_{2} < K_{1}}$ is shown in Fig. \ref{fig3} (a). It can be noted that MAE has a bean-shaped 3D surface, and the easy axis is aligned with the smallest $K_{i}$, $K_{3}$ in this scenario. If we consider the $xy$-plane MAE landscape, the in-plane easy axis lies aligned with $K_{2}$, which is the smaller one between $K_{1}$ and $K_{2}$. A similar argument is also valid for $xz$ and $yz$-planes. If $K_{6}$ becomes strong, it rotates the bean-shaped object and results in a tilted easy axis in a way that it is no longer aligned with any Cartesian coordinates (see Fig. \ref{fig3} (b)). 

To have insight into magnetic anisotropic features of FeXZ\(_2\) compounds, we calculated MAE as a function of Euler angles ($\theta$ and $\phi$), and we fitted Eq. \ref{eq2_mae} into results. The MAE values are listed in Table \ref{tab2} and the $K_{i}$ constants are listed in Table S2 in supplemental materials. The 3D and 2D MAE surfaces are shown in Fig. \ref{fig3}(c and d) for FeNbTe\(_2\) and FeTaTe\(_2\), respectively, and in Fig. S17, in the supplemental materials, for the rest of compounds. The magnitude of MAE is relatively high in all compounds, with the smallest 0.28 meV/Fe in FeNbS\(_2\) and the largest 1.57 meV/Fe in FeTaSe\(_2\). Compounds containing Ta show higher MAE, due to stronger spin orbit coupling in Ta compared to Nb. These results are comparable with perpendicular MAE in Fe\(_3\)GeTe\(_2\) ($\approx$1.22 meV/Fe), and CrI\(_3\) ($\approx$0.80 meV/Cr), \cite{webster2018strain, wang2019magnetic, ghosh2023unraveling}. On the other hand, $K_{6}$ is relatively strong and a canted easy axis appears along $xz$-plane in most of the compounds. The canting angle varies from 0.0$^{\circ}$ in FeTaSeTe to 41$^{\circ}$ in FeTaTe\(_2\). The synthesized FeNbTe\(_2\) shows a strong MAE of 0.47 meV/Fe and a canted angle of 13$^{\circ}$ in the $xz$-plane. The 3D and 2D MAE surfaces for FeXZ\(_2\) illustrate that anisotropy also exists in $xy$-and $yz$-planes. However, along these planes, the easy and hard-axis both are aligned with Cartesian coordinates. The strong and tilted magnetic anisotropy (TMA) is a rare and promising feature for applications in spin valves, spin transfer torque, and spin-orbit torque devices, where canted magnetism gives rise to multidirectional spin polarization and enables field-free magnetization switching \cite{zhao2023room, zhao2023coexistence}. TMA has been demonstrated in various systems, including Ta/CoFeB/MgO/Ta heterostructures~\cite{you2015switching}, Ru-substituted La$_{0.7}$Sr$_{0.3}$MnO$_3$ (Ru-LSMO) films under compressive strain~\cite{das2023tilted}, magnetic multilayers such as Gd/Co~\cite{kim2022field} and van der Waals heterostructures combining in-plane and out-of-plane easy magnetization materials like CrSBr/Fe$_3$GeTe$_2$~\cite{cham2024exchange}. However, the occurrence of TMA in freestanding and undoped monolayers is rare, making FeXZ\(_2\) monolayers promising candidates for spintronics applications.

To gain deeper insights into the trends observed in the MAE values, the atom-projected MAE is presented in Fig. \ref{fig3} (e). It is evident that the contribution from individual elements is not merely proportional to their atomic mass. Instead, atomic coupling appears to play a pivotal role in determining the observed values. Notably, the contribution from Fe atoms is significantly higher in the presence of Ta compared to Nb. This difference can be attributed to the \textit{d}-orbital hybridization between Fe and Ta, where, in the spin-up channel below the Fermi level, the overlap between Fe-Ta \textit{d}-states is greater than that of Fe-Nb \textit{d}-states, as illustrated in the projected density of states (PDOS) in Fig. S5 and S6 of the supplemental materials. Furthermore, it is observed that the contribution from Ta atoms in FeTaS\(_2\) is lower than in FeTaSe\(_2\) and FeTaTe\(_2\). This finding suggests that electronic hybridization between Ta and chalcogen atoms plays a crucial role in MAE, with the \textit{d-p} states overlapping in the presence of Se or Te being more pronounced compared to S atoms, due to more extended 4p orbitals of Se and 5p orbitals of Te.

\subsubsection*{Monte Carlo simulations}
To have an estimation of the magnetic transition temperature (T$_{C}$) in 2D FeXZ\(_2\) systems, we performed classical Monte Carlo (MC) simulations of temperature-dependent magnetic properties, considering localized spin moments, based on the following Heisenberg spin-Hamiltonian:

\begin{equation}
    H = -\sum_{i\neq j} J_{ij}\vec S_{i}{\cdot} \vec S_{j} -\sum_{i\neq j} D_{ij}\vec S_{i}{\times} \vec S_{j} - \sum_{i\neq j} K_{i}(S_{i}^{z})^{2}
    \label{H}
\end{equation}
where $J_{ij}$ represents isotropic symmetric exchange interactions between $i_{th}$ and $j_{th}$ sites, $D_{ij}$ is Dzyaloshinskii–Moriya interactions, and  $K_{i}$ is the single-ion anisotropy, and S$_{i}$ and S$_{j}$ are spin vectors. Fig. S18, in the supplemental materials, depicts the normalized susceptibility (normalized magnetization) vs. temperature for FeXZ\(_2\) compounds, where the peak of the susceptibility curve is the estimation of T$_{C}$. The T$_{C}$ values are listed in Table \ref{tab2}. Among FeXZ\(_2\) compounds, FeNbSe\(_2\) has the highest Curie temperature with T$_{C}$= 199 K, while lowest transition temperature appears in FeTaTe\(_2\) with T$_{C}$= 99 K. The synthesized FeNbTe\(_2\) exhibits a Curie temperature of 115~K, which is slightly higher than the experimentally observed value of 70~K \cite{wu2024tailoring, stepanova2024bulk}. This overestimation may arise from the absence of structural defects in the DFT model, as such defects are commonly present in experimental samples. Additionally, the intrinsic approximations of DFT calculations could further contribute to this difference. Another observation is that the transition temperatures in FeTaSe\(_2\) (T$_{C}$ = 155~K) and FeTaTe\(_2\) (T$_{C}$ = 99~K) are lower than that of FeTaS\(_2\) (T$_{C}$ = 195~K), while the magnetic anisotropy energy in FeTaS\(_2\) is significantly smaller compared to the other two compounds. This suggests that the exchange interaction term in the Hamiltonian of equation~\ref{H} dominates over the MAE term. Although the Curie temperatures (T$_{C}$) of these compounds are below room temperature, they remain higher than that of synthesized 2D CrI$_{3}$ (T$_{C}$ $\approx$ 45 K \cite{huang2017layer}) and are comparable to that of 2D Fe$_{3}$GeTe$_{2}$ (T$_{C}$ $\approx$ 130 K \cite{fei2018two}). Considering the strength of the first nearest-neighbor exchange interaction, $J_{ij}$, in these compounds ($J \approx 4 k_B T$), higher transition temperatures could be anticipated. However, it has been observed that the second nearest-neighbor $J_{ij}$ plays a dominant role in determining T$_{C}$. In other words, the first nearest-neighbor $J_{ij}$, with a coordination number of $n = 1$, couples only a single pair of adjacent Fe atoms. These isolated pairs, however, are interconnected through the second nearest-neighbor $J_{ij}$ (see Fig. \ref{fig2} (d)). Therefore, it is expected that the key to increasing T$_{C}$ lies in tuning the second nearest-neighbor $J_{ij}$, which could potentially be achieved through well-established techniques such as doping or strain engineering \cite{ghosh2024structural}.

\subsubsection*{DMI in Janus systems}
The Dzyaloshinskii-Moriya Interaction (DMI) is an antisymmetric exchange interaction in magnetic systems that arises due to spin-orbit coupling in the presence of broken inversion symmetry. This interaction plays a crucial role in stabilizing chiral spin textures, such as skyrmions, topologically protected quasi-particles characterized by a vortex-like or spiral spin configuration. Skyrmions have garnered significant attention in the field of spintronics due to their nanoscale dimensions, intrinsic stability, and efficient controllability, making them promising candidates for next-generation information storage and logic devices. In parallel, Janus structures represent a recently synthesized class of non-centrosymmetric two-dimensional (2D) materials. The first Janus system, MoSSe, was synthesized in 2017 through a selective replacement of the top-layer sulfur atoms with selenium in MoS$_2$ \cite{lu2017janus}. Following this breakthrough, other Janus systems, such as WSSe and PtSSe, have been successfully fabricated \cite{sant2020synthesis, trivedi2020room}. Given the recent experimental synthesis of FeNbTe$_2$ and ongoing advancements in Janus material engineering, we explore the potential for skyrmion formation in Janus FeNbSeTe and FeTaSeTe. To this end, we compute the exchange stiffness ($A$) and the spiralization matrix ($\mathbf{D}$), employing the following expressions,

\begin{equation}
A = \frac{1}{2} \sum_{j \neq i} J_{ij} R_{ij}^2 e^{-\mu R_{ij}} \quad
\label{A}
\end{equation}

\begin{equation}
D_{\alpha \beta} = \sum_{j \neq i} D_{ij}^{\alpha} R_{ij}^{\beta} e^{-\mu R_{ij}}
\label{D}
\end{equation}

where \( J_{ij} \) is the exchange interaction strength between sites \( i \) and \( j \), and \( R_{ij} \) is the distance between sites \( i \) and \( j \). The parameter \( \mu \) was varied between 2 and 4 and then extrapolated to \( \mu = 0 \) by fitting a third-order polynomial to improve numerical convergence with respect to the real-space cutoff (see Fig. S19 in the supplemental materials, which demonstrates exchange stiffness extrapolation for FeNbTe$_{2}$) \cite{borisov2024electronic, borisov2024tunable}.

The extracted results are presented in Table \ref{tab2}. Exchange stiffness (\( A \)) quantifies the tendency of magnetic moments to remain fully aligned. In other words, higher values of \( A \) correspond to a stronger ferromagnetic (FM) character and a higher Curie temperature (\textit{T$_{C}$}). It can be observed that FeXZ$_{2}$ and their Janus counterparts exhibit moderate exchange stiffness, which explains their relatively low magnetic transition temperatures. The highest exchange stiffness is found in FeTaS$_{2}$, with \( A = 421.7 \) meV\(\cdot\)Å$^{2}$, which also possesses one of the highest \textit{T$_{C}$} among all FeXZ$_{2}$ compounds. 

\begin{figure}[t!]
 \centering
 \includegraphics[width=\linewidth]{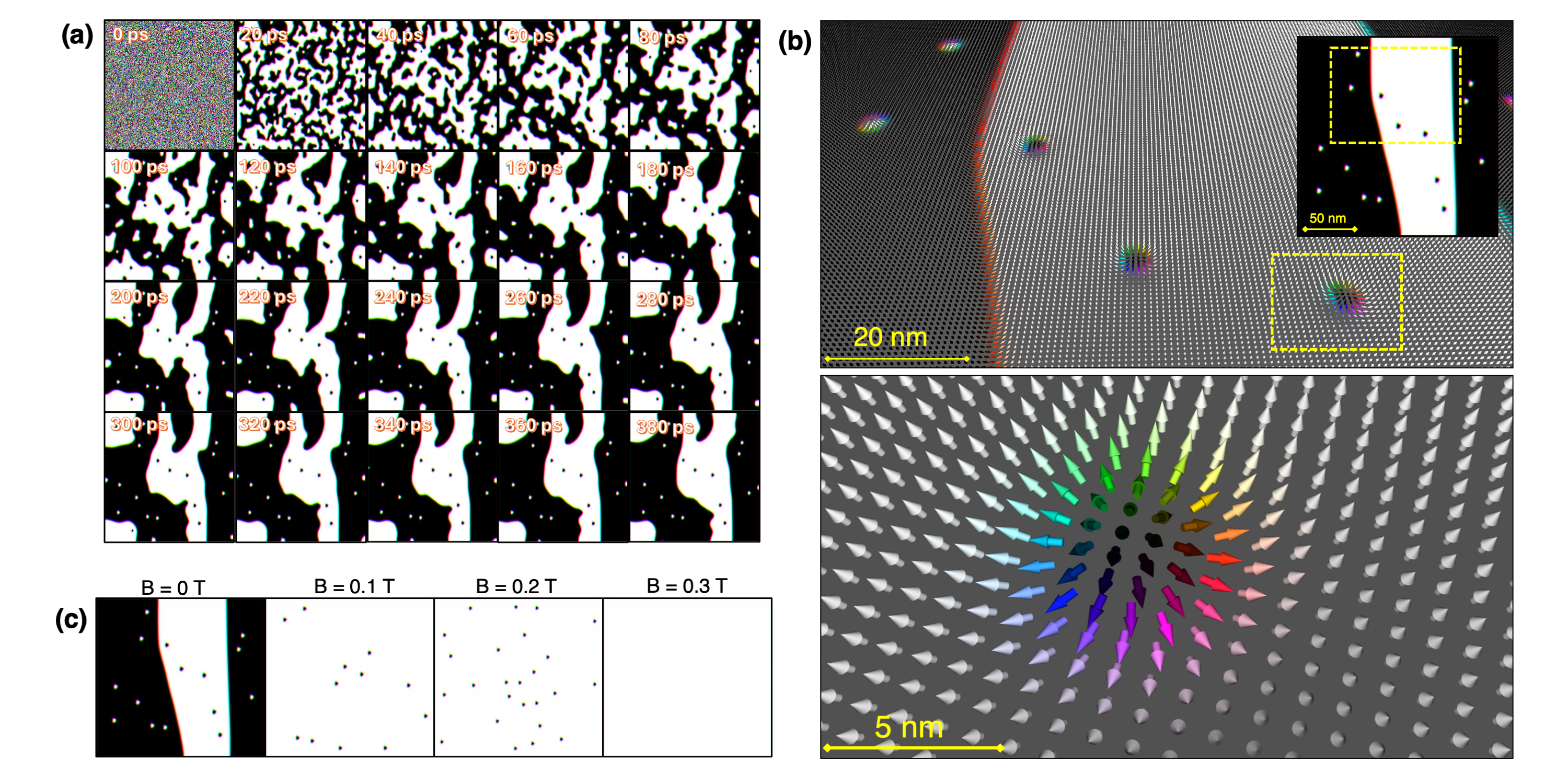}
  \caption{Micromagnetic Simulation of Magnetic Microstructure in Janus FeNbSeTe:
(a) Temporal evolution of the magnetic texture during the first 380 ps of the micromagnetic simulation, with snapshots taken every 20 ps, (b) Magnified view of skyrmionic spin texture, (c) Influence of external magnetic field on skyrmionic features.
}
  \label{fig4}
\end{figure}

On the other hand, the spiralization matrix satisfies Neumann’s symmetry principle, which states that the symmetry of the \( D \) matrix must incorporate the symmetry elements of the crystal’s point group. Among several types of Dzyaloshinskii-Moriya interaction (DMI), two of the most common are: (i) bulk DMI, arising from local inversion symmetry breaking within the atomic structure, and (ii) interfacial DMI, which occurs due to the absence of inversion symmetry at interfaces. In two-dimensional systems, interfacial DMI is typically dominant. Consequently, the spiralization matrix adopts a generic form as follows,

\[
\mathbf{D} =
\begin{bmatrix}
0 & D & 0 \\
-D & 0 & 0 \\
0 & 0 & 0
\end{bmatrix}
\]

where \( D \) represents the spiralization value, expressed in units of meV\(\cdot\)Å. The extracted spiralization values are also presented in Table \ref{tab2}. It is observed that in all centrosymmetric FeXZ$_{2}$ compounds, the spiralization value is zero. However, Janus systems exhibit finite spiralization values, with \( D = 20.4 \) meV\(\cdot\)Å for FeNbSeTe and \( D = 13.5 \) meV\(\cdot\)Å for FeTaSeTe. While nonzero spiralization values are a necessary condition for skyrmion formation, they are not sufficient on their own. The emergence of skyrmions depends on the intricate interplay between exchange stiffness, spiralization magnitude, magnetic anisotropy energy, and structural properties. To further explore this possibility, we performed micromagnetic simulations for FeNbSeTe and FeTaSeTe, where magnetization dynamics was modeled using the Landau-Lifshitz-Gilbert (LLG) equation, given by,

\begin{equation}
\frac{\partial \mathbf{m}}{\partial t} = -\gamma \mathbf{m} \times \mathbf{H_{\text{eff}}} + \alpha \mathbf{m} \times \frac{\partial \mathbf{m}}{\partial t}
\label{LLG}
\end{equation}

where $\gamma$ is the gyromagnetic ratio, $\mathbf{H_{\text{eff}}}$ is the effective magnetic field, which includes contributions from exchange, anisotropy, Dzyaloshinskii-Moriya interactions, and the external magnetic field, and $\alpha$ (=0.1) is the damping parameter. The material parameters used in micromagnetic simulations were obtained from \textit{ab initio} calculations and converted into their continuous forms, including spin stiffness (\( A \)), spiralization value (\( D \)), uniaxial anisotropy constant (\( K_{u1} \)), and magnetic moment per unit cell ($M_{s}$). It was found that in FeTaSeTe, the magnetic anisotropy energy (MAE) is sufficiently strong to prevent the formation of any chiral magnetic textures. At low temperatures, the magnetic moments remain aligned along the easy axis. As the temperature increases, thermal fluctuations eventually overcome the dominance of MAE, leading to a ferromagnetic-to-paramagnetic transition. For FeNbSeTe, however, a different scenario emerges. In this system, the \( D/A \) ratio is found to be 0.08, which is favorable for skyrmion formation, while the MAE remains moderate. The evolution of the magnetic microstructure as a function of time for FeNbSeTe is depicted in Fig. \ref{fig4}(a). It is observed that starting from a fully random magnetic configuration, the microstructure evolves into a ferromagnetic texture with large domains, where black and white regions correspond to spin-down and spin-up states, respectively. Additionally, within the ferromagnetic domains, small black and white dots appear. Fig. \ref{fig4}(b) provides a magnified view of the spin texture of these dots. It is evident that these spin textures resemble Néel-type skyrmions, with an approximate diameter of 8–9 nm. To confirm the topological nature of these structures, we computed the topological winding number using the following equation,

\begin{equation}
N_{\text{sk}} = \frac{1}{4\pi} \int \mathbf{m} \cdot 
\left( \frac{\partial \mathbf{m}}{\partial x} \times \frac{\partial \mathbf{m}}{\partial y} \right) 
\, dx \, dy.
\label{winding}
\end{equation}

A topological winding number close to \( N_{\text{sk}} = 1 \) was found for all the observed dots in the system (Due to inaccuracies associated with finite difference derivatives in Eq.~(10), the winding number does not yield an exact integer \cite{kim2020quantifying}.) This confirms the topological nature of the spin texture and indicates that these structures are Néel-type skyrmions. The effect of an external magnetic field on the skyrmionic features of FeNbSeTe is illustrated in Fig. \ref{fig4}(c). It can be observed that at \( B = 0.1 \) T and \( B = 0.2 \) T, the magnetic domains vanish, while skyrmions persist within a single-domain ferromagnetic background. However, increasing the external field to \( B = 0.3 \) T leads to the complete destruction of skyrmions. The stabilization and manipulation of skyrmions in the presence of an external magnetic field has been reported in numerous magnetic systems. However, zero-field skyrmions are highly sought after, as they are particularly desirable for device applications. Zero-field Néel-type skyrmions are reported in multiferroic lacunar spinel GaV$_{4}$S$_{8}$ \cite{borisov2024dzyaloshinskii}. Also, they have been observed in FeGe thin films epitaxially grown on Si(111), where skyrmion formation was confirmed through topological Hall effect measurements~\cite{gallagher2017robust}. Furthermore, zero-field skyrmions have been experimentally observed in PtMnGa~\cite{srivastava2020observation} and Cr$_{1+0.3}$Te$_2$~\cite{saha2022observation}. The potential for skyrmion formation in Janus FeNbSeTe in the absence of an external magnetic field places this material among the rare family of zero-field skyrmionic 2D materials. This characteristic makes FeNbSeTe a promising candidate for further experimental synthesis and investigation.

In conclusion, a systematic study of structure, stability, electronic and magnetic properties of recently synthesized monoclinic FeNbTe\(_2\) and its family (FeXZ\(_2\)) in the monolayer regime is presented. All 2D compounds were found to be energetically, dynamically, thermally, and mechanically stable. Evolutionary structural search indicates that FeNbTe\(_2\) retains its monoclinic symmetry even in the monolayer form. The anaysis of magnetic exchange interaction parameters reveal that both direct and indirect exchange mechanisms are active in these compounds, where the first nearest neighbor interaction is mainly direct, and the second nearest interaction and beyond are indirect, mediated by the chalcogen atoms. Very interestingly, it was found that FeXZ\(_2\) compounds offer a canted out-of-plane magnetic anisotropy, which is a rare feature, sought after in spintronic applications. The magnitude of MAE was found to be relatively high, varying between 0.28 to 1.57 meV/Fe. Magnetic transition temperature T$_{C}$ obtained from Monte Carlo simulations showed that that T$_{C}$ is lower than room temperature for all compounds but higher than the cryogenic temperature for most of them. Furthermore, the Janus counterparts (FeNbSeTe and FeTaSeTe) were studied. It was found that although pristine FeXZ\(_2\) systems do not show a strong spiralization value due to centrosymmetry, their non-centrosymmetric Janus counterparts show a strong spiralization value. Micromagnetic simulations were performed for Janus systems, and it was found that FeNbSeTe can host Néel type skyrmions even in the absence of an external magnetic field. Considering the successful synthesis of FeNbTe\(_2\), and recent advances in growing Janus crystals, future experimental studies of FeNbSeTe can be done in the quest for zero-field skyrmions. 

\section*{Methods}

First-principles structural optimization, stability analysis, magnetic ground state, and magneto-crystalline anisotropy energy were conducted in the framework of density functional theory (DFT) \cite{kohn1965self,hohenberg1964inhomogeneous} using the Vienna \textit{Ab initio} Simulation Package (VASP) \cite{kresse1993ab,kresse1994ab,kresse1996efficiency,kresse1996efficient}. The exchange-correlation potential was treated using the generalized gradient approximation (GGA) functional in conjunction with the Perdew, Burke, and Ernzerhof (PBE) method \cite{perdew1996generalized}. The projector augmented wave method \cite{Blochl1994projector} was applied. A plane-wave basis set with a kinetic cutoff energy of 500 eV was used to expand the electronic wave function, and a vacuum space of at least 20 \AA~ was inserted along the $c$-axis to prevent unrealistic interactions between periodic images. During structural optimization, the maximum force on each atom was less than  5$\times$10$^{-3}$ eV/\AA~. A~Gaussian smearing factor of 0.05 was taken into account. Brillouin zone (BZ) integration was performed by a $\Gamma$-centered $7\times9\times1$  uniform $k$-point grid for monolayers. In order to determine the net charge transfer between constituent atoms, the Bader technique was employed \cite{henkelman2006fast}. To ensure the dynamical stability of the systems, the phonon dispersions were calculated, using the finite-displacement approach, implemented in the PHONOPY \cite{togo2015first} package. To analyze the thermal stability, \textit{ab initio}  molecular dynamics (AIMD) simulations were conducted, using a microcanonical ensemble (NVE), at constant temperatures of T = 500 K for a simulation period of 6 ps with 1 fs time steps. A $2\times3\times1$ supercell was used for phonon dispersions and AIMD. Evolutionary crystal structure prediction was performed using the USPEX code \cite{oganov2006crystal}. For the calculation of magnetic anisotropy energy (MAE), a $15\times19\times1$ $k$-point grid was utilized. For the MAE calculations, the spin–orbit coupling (SOC) was taken into account. Heisenberg exchange interaction, DMI, MAE, and atom-resolved MAE were calculated via the QuantumATK-Synopsys package version U-2022, employing an LCAO basis set, the "PseudoDojo" pseudopotential \cite{qatk1, qatk2}, a density mesh cutoff of 120 Hartree, and a $15\times19\times1$ $k$-point grid. The projection into localized Wannier Functions (WFs) was carried out using the Wannier90 package \cite{mostofi2008wannier90}, via VASP to Wannier90 interface. The WFs basis set comprised five d-orbitals of transition metal and three p-orbitals of chalcogen (S, Se, Te) atoms. The tight-binding hopping parameters were extracted from WFs. The TB2J \cite{he2021tb2j} package was utilized to calculate orbital resolved isotropic exchange interactions via Liechtenstein, Katsnelson, Antropov, and Gubanov (LKAG) \cite{liechtenstein1987local} formalism. The extracted exchange interactions were implemented in a Heisenberg Hamiltonian to calculate the magnetic ordering temperature by performing classical Monte Carlo (MC) simulations via UppASD code \cite{eriksson2017atomistic}. To achieve properly averaged properties, five ensembles within a supercell of $60\times80\times1$ were modeled, assuming periodic boundary conditions. Micromagnetic simulations were performed using MuMax3\cite{vansteenkiste2014,mulkers2017}. The unit cell dimensions were obtained from relaxed structures in \textit{ab initio} calculations and were used to form a grid of $256 \times 256 \times 1$ cells. The Gilbert damping parameter was set to $\alpha = 0.10$ as proposed in a previous study \cite{joos2023tutorial}, and uniaxial anisotropy was applied. The simulation time was 1 ns for each set of parameters, ensuring the system would reach an equilibrium state. The values for spin stiffness ($A$), DMI ($D$), uniaxial anisotropy constant ($K_{u1}$), and magnetic moment per unit cell ($\mu$) were taken from \textit{ab initio} calculations and converted into the continuous material parameters used in micromagnetic simulations. The system was subjected to an external magnetic field applied perpendicular to the atomic plane, with field magnitudes ranging from 0~T to 0.5~T in steps of 50~mT for the initial calculation. To achieve more detailed data for subsequent analysis, the interval was decreased to 5~mT.

\bibliography{references}

\section*{Acknowledgements}

B.S. acknowledges financial support from Swedish Research Council (grant no. 2022-04309 and grant No. 2018-07082). The computations were enabled by resources provided by the National Academic Infrastructure for Supercomputing in Sweden (NAISS) at UPPMAX (NAISS 2024/5-258) and at NSC and PDC (NAISS 2024/3-40) partially funded by the Swedish Research Council through grant agreement no. 2022-06725. B.S. and S. E. also acknowledge EuroHPC for awarding us access to EHPC-DEV-2024D03-043 hosted by LUMI in Finland, and EU2023D11-039 hosted by Karolina at IT4Innovations, Czech Republic. Artificial intelligence was used to improve the language and readability of this work.

\section*{Author contributions statement}

S.E.: Investigation (lead); Data curation (lead); Formal analysis (lead); Visualization (lead); Writing - original draft (lead). N.M. : Data curation (supporting); Formal analysis (supporting); Writing - review \& editing (supporting). A.M. : Formal analysis (supporting). V.B. : Formal analysis (supporting). O.E. : Formal analysis (supporting). B.S. : Resources (lead); Formal analysis (supporting); Methodology (lead); Writing - review \& editing (lead); Funding acquisition (lead); Supervision (lead). All authors reviewed the manuscript.

\section*{Data availability}
All significant results are presented in the main manuscript and supplementary materials. Additional data supporting this study are available from the corresponding author upon reasonable request.

\section*{Code availability}
For our DFT calculations, we employed VASP, QuantumATK, Wannier90, TB2J, USPEX, and Phonopy. Monte Carlo and micromagnetic simulations were conducted using UppASD and Mumax. All these software packages are either commercially available or open-source.

\section*{Conflict of interest}
The authors declare that they have no conflicts of interest to disclose.


\end{document}